\documentclass[pra,aps,amsfonts,amsmath,superscriptaddress,twocolumn,showpacs,floatfix]{revtex4-1}
\usepackage{graphicx} 

\begin{document}

\title{Incompressibility of finite fermionic systems: 
stable and exotic atomic 
nuclei}

\author{E. Khan}
\affiliation{Institut de Physique Nucl\'eaire, Universit\'e Paris-Sud, IN2P3-CNRS, F-91406 Orsay Cedex, France}

\author{N. Paar}
\author{D. Vretenar}
\affiliation{Physics Department, Faculty of Science, University of
Zagreb, 10000 Zagreb, Croatia}

\author{Li-Gang Cao}
\affiliation{Institute of Modern Physics, Chinese Academy of Science,
Lanzhou 730000, P. R. China}
\author{H. Sagawa}
\affiliation{Center for Mathematics and Physics, University of Aizu, Aizu-Wakamatsu, 965-8580 Fukushima, Japan}
\affiliation{RIKEN Nishina Center, Wako 351-0198, Japan}

\author{G. Col\`o}
\affiliation{Dipartimento di Fisica, Universit\`a degli Studi and 
INFN Sez. di Milano, Via Celoria 16, 20133 Milano, Italy}

\begin{abstract}
The incompressibility of finite fermionic systems is investigated using 
analytical approaches and microscopic models. The incompressibility 
of a system is directly linked to the zero-point kinetic energy of 
constituent fermions, and this is a universal feature of fermionic 
systems.  In the case of atomic nuclei, this implies a 
constant value of the incompressibility in medium-heavy and heavy 
nuclei. The evolution of nuclear incompressibility along Sn and 
Pb isotopic chains is analyzed using global microscopic models, 
based on both non-relativistic and relativistic energy functionals. 
The result is an almost constant incompressibility in stable 
nuclei and systems not far from stability, and a steep decrease in 
nuclei with pronounced neutron excess, caused by the emergence of 
a soft monopole mode in neutron-rich nuclei.
\end{abstract}


\date{\today}

\maketitle

\section{Introduction \label{intro}}

Recent advances in experimental and theoretical investigations of 
nuclear incompressibility have also enabled studies of the evolution
of incompressibility in isotopic chains
\cite{tli07,tli10,gar09,jli08,kha09,ves12}. As for many other
structure phenomena, for instance the evolution of magic numbers
\cite{sor08}, a study of the isotopic dependence might provide a more
general and deeper insight into the inner workings.  It has recently
been inferred that measurement of isoscalar giant monopole resonances
(ISGMR) may probe nuclear incompressibility in a range of low
nucleonic densities, rather than exactly at saturation density
\cite{kha12}. Experiments on isotopic chains \cite{tli10,gar09},
possibly also extended to exotic nuclei \cite{mon08,van10}, are
therefore important for constraining the density dependence of the
incompressibility. Investigations of isotopic chains also enable
studies of pairing and shell effects on incompressibility. It has been
shown that one cannot neglect the effect of shell closure on nuclear
incompressibility, and pairing correlations affect the position of the
GMR excitation energy \cite{jli08,kha09}. The effect of shell closure
(and therefore magicity) on the incompressibility has been discussed
in Ref.~\cite{kha09b}. However, there is a general consensus that
pairing effects alone are not large enough to explain the available
data \cite{ves12,lig12,gar12}. Early microscopic studies of the 
nuclear incompressibility were carried out using Skyrme effective
forces. Later, with the inclusion of relativistic mean field models,  
detailed comparisons between different model predictions led to
the conclusion that the nuclear matter incompressibility extracted from e.g.
$^{208}$Pb lies in the range $\approx$ 240$\pm$20 MeV (cf. \cite{sch06}
and references therein). The remaining uncertainty reflects our
incomplete understanding of the specific effects related to finite
nuclei, and further justifies the need of studying extended isotopic
chains.

The theoretical framework of energy density functionals (EDF)
\cite{ben03,vre05} is the tool of choice for a quantitative analysis
of these phenomena. Preliminary studies, undertaken in the 1980's,
analyzed the dependence of the nuclear incompressibility on mass
number, but only four doubly magic nuclei were considered and 
it was suggested that the nuclear
incompressibility is rather constant for heavy nuclei
\cite{bla80,boh79}. Another aspect is that the occurrence of 
low-energy monopole strength in exotic nuclei has recently been predicted 
\cite{ham98,cap09,kha11}, and this might affect the nuclear
incompressibility. More generally, the present study of extended
isotopic chains could also be useful for the understanding of
incompressibility in other fermionic systems, or for astrophysical
processes such as core collapse supernovae. The use of both analytical
and fully microscopic approaches will provide complementary information 
about the incompressibility in both stable and exotic nuclei.

Section \ref{analytic} defines the analytical relations between the
incompressibility and other relevant nuclear quantities. Section \ref{micro} compares
the analytical results with the microscopic ones obtained using 
the EDF framework. A brief summary and conclusions  are given in section \ref{conclusion}.

\section{Incompressibility of a Finite Fermionic System \label{analytic}}

In this section we review the relations between the nuclear 
incompressibility K$_A$ and the nucleonic kinetic energy in the following 
analytical models: infinite nuclear matter, 
Fermi gas, and the harmonic oscillator. For instance, in the free
Fermi gas approach the incompressibility of the system is directly 
linked to the Fermi energy \cite{lan67}. 

\subsection{Definitions}

The incompressibility of a fermionic system composed
of A particles is defined as the second derivative (curvature) of its energy 
with respect to the mean square radius at the minimum \cite{bla80}:

\begin{equation}
K_A\equiv \left. 4\langle r^2 \rangle\frac{d^2 E/A}{d\langle r^2
\rangle^2}\right|_{g.s.} \; ,
\label{eq:kaone}
\end{equation}
where the label $g.s.$ indicates that the quantities are evaluated in the ground state.
For the isoscalar monopole transition 
operator
\begin{equation}
\hat{Q}=\sum_{i=1}^A r_i^2 \; ,
\label{mon_operator}
\end{equation}
the dielectric theorem \cite{boh79,bla80,cap09} yields
\begin{equation}
K_A=\frac{2 A\langle r^2\rangle^2_{g.s.}}{m_{-1}}\; ,
\label{eq:katwo}
\end{equation}
where $m_{-1}$ is the inverse energy weighted sum rule, proportional to the
polarizability of the system. 

The energy of the isoscalar GMR can be defined by the following relation \cite{bla80,boh79}: 
\begin{equation}
E_{\rm GMR}=\sqrt{\frac{m_1}{m_{-1}}}.
\label{eq:egsum}
\end{equation} 
The moment $m_1$ is evaluated by the double commutator using the 
Thouless theorem \cite{tho61}:
\begin{equation}
m_1=\frac{2\hbar^2A}{m} \langle r^2 \rangle_{g.s.} \;,
\label{eq:m1}
\end{equation}
where $m$ is the fermion mass, that is, $mc^2\approx$ 938 MeV for the nucleon.
Eq. (\ref{eq:katwo}), (\ref{eq:egsum}) and (\ref{eq:m1}) lead to 
\cite{bla80}:
\begin{equation}
E_{\rm GMR}=\sqrt{\frac{\hbar^2K_A}{m \langle r^2 \rangle}}
\label{eq:ka}
\end{equation}

%It should be noted that the present definition (\ref{eq:kaone}) of K$_A$
%imposes the use of the polarisability in the evaluation of the GMR
%energy (\ref{eq:egsum}). The incompressibility K$_A$ of a finite
%system can be considered using the above equations on a sound footing: 
%it is a property of the system under consideration.

\subsection{An analytical study}

An analytical derivation of the nuclear incompressibility
K$_A$ in the nuclear matter, the Fermi gas,
and the harmonic oscillator models, will provide a complementary
insight to the microscopic results of the next section.

\subsubsection{Nuclear matter and the Fermi gas}

Let us first consider infinite homogeneous nuclear matter. 
The nuclear matter incompressibility at 
saturation $K_\infty$ is understood fairly well, whereas the behavior of the, 
more general, density-dependent incompressibility has not been studied much.
By defining the general 
density-dependent incompressibility K($\rho$) \cite{fet71,kha12}
%--------------------------------------------------------------------------------------------
\begin{equation}
K(\rho)\equiv\frac{9}{\rho \chi(\rho)}=9\rho^2\frac{\partial^2
E(\rho)/A}{\partial
\rho^2} + \frac{18}{\rho} P(\rho)\; ,
\label{eq:krho}
\end{equation}
%--------------------------------------------------------------------------------------------
and neglecting inter-nucleon interactions, 
one finds \cite{bla80,lan67}
%--------------------------------------------------------------------------------------------
\begin{equation}
K(\rho)=6\epsilon_F=\frac{6\hbar^2}{2m}\left(\frac{3\pi^2}{2}\right)^{2/3}\rho^{2/3}\; .
\label{eq:rinf}
\end{equation}
%--------------------------------------------------------------------------------------------
This equation exhibits the density dependence of the
incompressibility, known to play an important role in the relation
between the measurement of GMRs and the equation of state
\cite{kha12}: in the case of non-interacting nuclear matter the
density-dependent incompressibility is proportional to $\rho^{2/3}$.
At saturation density ($\rho_0$=0.16 fm$^{-3}$) Eq.~(\ref{eq:rinf})
yields $K(\rho_0)$=220 MeV, and at the so-called crossing density
\cite{kha12} ($\rho_c$=0.1 fm$^{-3}$) one gets $K(\rho_c)$=160 MeV.
This shows that the density dependence of the incompressibility of
non-interacting nuclear matter presents a good approximation at
saturation density but not at the crossing density, for which
$K_c\simeq$40 MeV is the expected value \cite{kha12}. This can be
understood because at the crossing density interaction effects are
expected to be more important, whereas at saturation density the
homogeneous nucleonic system can simply be modeled by non-interacting
hard spheres.  Considering an average spacing $r_0$ among the
constituents, one gets for the homogeneous density of the system:
%--------------------------------------------------------------------------------------------
\begin{equation}
\rho=\left(\frac{4}{3}\pi r_0^3\right)^{-1} \;.
\label{eq:r0}
\end{equation}
%--------------------------------------------------------------------------------------------
In the case of nucleons, $r_0\approx$1.2 fm yields $\rho_0\approx$ 0.15 fm$^{-3}$, 
and this shows the validity of the model in the vicinity of saturation density. 

Approximating the nucleus by a free Fermi gas of nucleons, we study  a spherical
fermionic A-body system with a typical equilibrium inter-nucleon distance
$r_0$. The radius of the system is then $R$=$r_0A^{1/3}$. To establish a link between the
incompressibility $K_A$ and the zero-point energy, one inserts  
Eq. (\ref{eq:r0}) in Eq. (\ref{eq:rinf}), and this leads to
%--------------------------------------------------------------------------------------------
\begin{equation}
K_A=\frac{3}{4}(9\pi)^{2/3}T_0 \approx  7 T_0 \;, 
\label{eq:rinf2}
\end{equation}
%--------------------------------------------------------------------------------------------
where $T_0$ is the zero-point kinetic energy (ZPE) of a fermion in the
system \cite{mot96}: 
%--------------------------------------------------------------------------------------------
\begin{equation}
T_0\equiv\frac{\hbar^2}{mr_0^2} \;.
\label{eq:t0}
\end{equation}

The ZPE,  first discussed by Mottelson in Ref.~\cite{mot96}, depends only on 
the characteristics of the constituent fermions, 
but not on the system itself. For nucleons
$T_0 \approx$ 30 MeV, and this yields $K_A \approx$ 210 MeV. It should be noted
that Blaizot also derived a relation between K$_A$ and the average
kinetic energy \cite{bla80}. However, the average kinetic energy
depends on the model and the system considered. Here the
incompressibility is linked to the ZPE, which does not depend on any
model nor on the specific system.  Eq.~(\ref{eq:rinf2})
shows that the larger the ZPE, the system becomes more difficult to compress: 
quantum localization of individual nucleons determines the
nuclear incompressibility. As a consequence, $K_A$ should be
independent of A because the incompressibility is directly linked to
the nucleonic ZPE (provided, of course, that the surface to volume
ratio is small enough).

\subsubsection{Harmonic oscillator}

In the harmonic oscillator (HO) approximation the $L=0$ 
radial compression mode (GMR) can be represented by 
a 2$\hbar\omega$ mode \cite{VdW}, where $\hbar\omega$ 
is 
%\cite{rin80}
%--------------------------------------------------------------------------------------------
\begin{equation}
\hbar\omega \approx
\frac{5}{4}\left(\frac{3}{2}\right)^{1/3}\frac{\hbar^2}{mr_0^2}A^{-1/3} \; .
\label{eq:ho}
\end{equation}
%--------------------------------------------------------------------------------------------
%It should be noted that this 2$\hbar\omega$ approximation of the GMR
%energy does not consider the effect of the residual interaction.

Assuming spherical density, $\langle r^2 \rangle$ in Eq.~(\ref{eq:ka}) can be
approximated by $3R^2$/5 where $R$ is the nuclear radius, %\cite{rin80}, 
and the GMR energy becomes:
%--------------------------------------------------------------------------------------------
\begin{equation}
E_{\rm GMR}=\sqrt{\frac{5\hbar^2K_A}{3mr_0^2}}A^{-1/3} \;.
\label{eq:ka2}
\end{equation}
%--------------------------------------------------------------------------------------------
In the HO approximation  $E_{\rm GMR}$=2$\hbar\omega$, and from Eqs.~(\ref{eq:ho}) and (\ref{eq:ka2}) one obtains:
%--------------------------------------------------------------------------------------------
\begin{equation}
K_A \approx
\frac{15}{4}\left(\frac{3}{2}\right)^{2/3}\frac{\hbar^2}{mr_0^2} \approx
5 T_0 \;.
\label{eq:katop}
\end{equation}
%--------------------------------------------------------------------------------------------

Therefore, in the HO approximation  
the compression modulus is also found to be constant and
directly related to the zero-point kinetic energy (ZPE). This again means
that the higher the ZPE, the harder it becomes to
compress the system. For nuclei $K_A$ is expected to be a constant 
in the lowest-order approximation, depending only on the nucleon mass 
and the typical inter-nucleon distance. Eq.~(\ref{eq:katop}) yields
$K_A$ $\simeq$ 140 MeV, and one expects that the HO value is more accurate 
than the one obtained in the Fermi gas model. This is  because the HO potential 
presents a much better first-order approximation to
the nuclear mean-field potential. 

If independent of A, however, the value $K_A$ $\approx$ 140 MeV does not 
correspond to the limit of infinite nuclear matter when $A \rightarrow \infty$, 
that is, $K_\infty$=220 MeV. The 
reason is that the HO value has been derived under the assumption that 
the GMR is a 2$\hbar\omega$ mode, but this is not valid any more for infinite
nuclear matter. In that case, it is the incompressibility of the Fermi
gas that should be used, yielding $K_A$=210 MeV as described above.
This difference is also confirmed by the fact that: i) $K_A \approx 140$ MeV is predicted 
by microscopic calculations in finite nuclei \cite{kha09,ves12}, and ii) 
$K_\infty \geq 200$ MeV is obtained by microscopic 
extrapolation of GMR
measurements to infinite nuclear matter \cite{kha12}.

\section{Microscopic calculation of the nuclear incompressibility
in isotopic chains  \label{micro}}

The first fully microscopic studies of the behavior of K$_A$ as a
function of the number of nucleons were performed in the 80's
\cite{bla80,boh79}.  At that time only four spherical magic nuclei
were considered: $^{16}$O, $^{40}$Ca, $^{90}$Zr and $^{208}$Pb. It was
inferred that $K_A$ is not very sensitive to A, although a general
conclusion was hindered by the small number of nuclei included in the
analysis. One should also emphasize that light nuclei are unlikely to
fit into the general picture because of the large surface to volume
ratio. In light nuclei the experimental ISGMR strength distribution 
displays pronounced fragmentation, and it is hard to extract a 
well-defined value for the monopole energy.

To investigate the evolution of nuclear incompressibility along isotopic chains, 
it is necessary to use global, EDF-based models, that also take into account pairing
effects. The constrained-HFB (CHFB) and the quasiparticle random-phase
approximation (QRPA) \cite{paa07} are employed in the present study. For the 
latter, both relativistic and non-relativistic implementations will be used to 
reduce possible model dependencies. Our aim is to verify, in a microscopic 
calculation, the constant behavior of $K_A$ predicted by the analytical model 
described in the previous section.

\subsection{Formalism}

Microscopic EDF-based calculations of isoscalar GMRs 
are usually performed using the CHFB or the QRPA frameworks \cite{paa07}. 
In this study we calculate the GMR excitation energies 
using the CHFB approach based on a Skyrme EDF,  and QRPA calculations 
are carried out  using both Skyrme and relativistic functionals.

\subsubsection{Constrained-HFB}

The extension of the constrained HF method
\cite{boh79} to include pairing correlations, that is, the CHFB 
approach has been detailed in Ref.
\cite{cap09}, and applied also in \cite{kha09}. In the CHFB the
energy of the GMR is computed using Eq. (\ref{eq:egsum}), with the $m_1$
moment determined by the HFB r.m.s. radius (Eq. \ref{eq:m1}). To evaluate the 
$m_{-1}$ moment, the constraint associated with the isoscalar monopole operator 
is added to the HFB Hamiltonian
%---------------------------------------------------------------------------------------------------------------
\begin{equation}
\hat{H}_{constr.}=\hat{H}+\lambda\hat{Q} \;.
\end{equation}
%---------------------------------------------------------------------------------------------------------------
The $m_{-1}$ moment is obtained from the derivative of the expectation value of 
the monopole operator in the CHFB eigenstate $\vert\lambda\rangle$,
using the dielectric theorem
%---------------------------------------------------------------------------------------------------------------
\begin{equation}
m_{-1}=-\frac{1}{2}\left[\frac{\partial}{\partial\lambda}\langle\lambda|
\hat{Q}|\lambda\rangle\right]_{\lambda=0} \;.
\end{equation}
%---------------------------------------------------------------------------------------------------------------

In the present calculation we employ the Skyrme functional SLy5 \cite{cha98}, 
together with the density-dependent delta-pairing interaction
%---------------------------------------------------------------------------------------------------------------
\begin{equation}\label{eq:vpair}
V_{pair}=V_0\left[1-\eta\left(\frac{\rho(r)}{\rho_0}\right)^\alpha\right]
\delta\left({\bf r_1}-{\bf r_2}\right)\;, 
\end{equation}
%---------------------------------------------------------------------------------------------------------------
with $V_0$=-285 MeV fm$^{3}$, 
$\eta$=0.35, $\alpha$=1 and $\rho_0$=0.16 fm$^{-3}$. This
pairing interaction has been designed to reproduce the experimental
two-neutron separation energies for the even-even isotopes $^{114-134}$Sn
\cite{kha09c}. The energy cutoff (quasiparticle energy)
 is $E_{max}=60$ MeV, and $j_{max} = 15/2$. 

\subsubsection{Quasiparticle Random Phase Approximation}

For completeness we briefly outline the details of the QRPA
calculations. The Skyrme HF-BCS equations for the ground state are
solved in coordinate space. The radial mesh extends to 18 fm, with a
spacing of 0.1 fm. For all nuclei analyzed in the present study we
have verified the stability of the results with the variation of  the size of the mesh.
Pairing correlations are taken into account by a contact
volume-pairing interaction, with the strength $V_0=-218 (-265)$ MeV
fm$^{3}$ for Sn (Pb) isotopes. The pairing strength is adjusted to reproduce 
the empirical pairing gaps extracted from data on odd-even
mass differences (five-point formula). The single-particle
continuum is discretized by placing nuclei in a spherical box of
radius 18 fm. For each set of spherical single-particle quantum
numbers ($l$, $j$), the QRPA model space includes unoccupied states up
to the maximum number of nodes given by $n_{max}$ = $n_{\rm last}$+12,
where $n_{\rm last}$ is the number of nodes for the last occupied
state with a given ($l$, $j$). The convergence of the results has been
verified by calculating the strength function of the ISGMR. For a
state with good angular momentum and parity $J^\pi$, the QRPA matrix
equation reads
%---------------------------------------------------------------------------------------------------------------
\begin{eqnarray}\label{RPA} \left( \begin{array}{cc}
  A & B \\
B^* & A^* \end{array}  \right) \left( \begin{array}{c}
X^n \\
Y^n  \end{array} \right) =\hbar\omega_n \left( \begin{array}{cc}
1 & 0\\
0 & -1 \end{array}  \right) \left( \begin{array}{c}
X^n \\
Y^n  \end{array} \right), \end{eqnarray} 
%---------------------------------------------------------------------------------------------------------------
where $\hbar \omega_n$ is the energy
of the $n$-th QRPA state. X$^n$ and Y$^n$ are the corresponding
forward and backward amplitudes, respectively. The explicit expressions 
for the matrices A and B are given, for instance, in Refs.~ \cite{rin80,colo13}. 
The particle-hole ($p-h$) matrix elements are computed from the Skyrme energy 
density functional including all terms, that is, also the two-body spin-orbit
and two-body Coulomb interactions.

The resulting moments of the QRPA strength distribution are calculated as %--------------------------------------------------------------------------------------------------------------- 
\begin{equation}
m_k=\int E^kS(E)dE,
\end{equation}
%--------------------------------------------------------------------------------------------------------------- 
where $S(E)=\sum_n|\langle 0|\hat{Q}|n\rangle|^2\delta(E-E_n)$ is
the strength function associated with the monopole operator 
defined in Eq.~(\ref{mon_operator}).

In the present analysis we also employ the QRPA based on relativistic
nuclear energy density functionals. More details about this
framework are given in Refs.~\cite{VALR.05,NVFR.02,PRNV.03}. It is
realized in terms of the fully self-consistent RQRPA based on the
relativistic Hartree-Bogoliubov model (RHB). The RQRPA equations are
formulated in the canonical single-nucleon basis of the RHB model, by
employing effective Lagrangians with density-dependent
meson-nucleon couplings~\cite{NVFR.02} in the particle-hole channel,
and the finite-range Gogny interaction~\cite{BGG.91} in the pairing
channel. The present calculation of the isoscalar monopole response is
carried out using the density-dependent meson-exchange effective 
Lagrangian DD-ME2 \cite{LNVR.05}.

\subsection{Results and interpretation}

\begin{figure}[tb]
\begin{center}
\scalebox{0.35}{\includegraphics{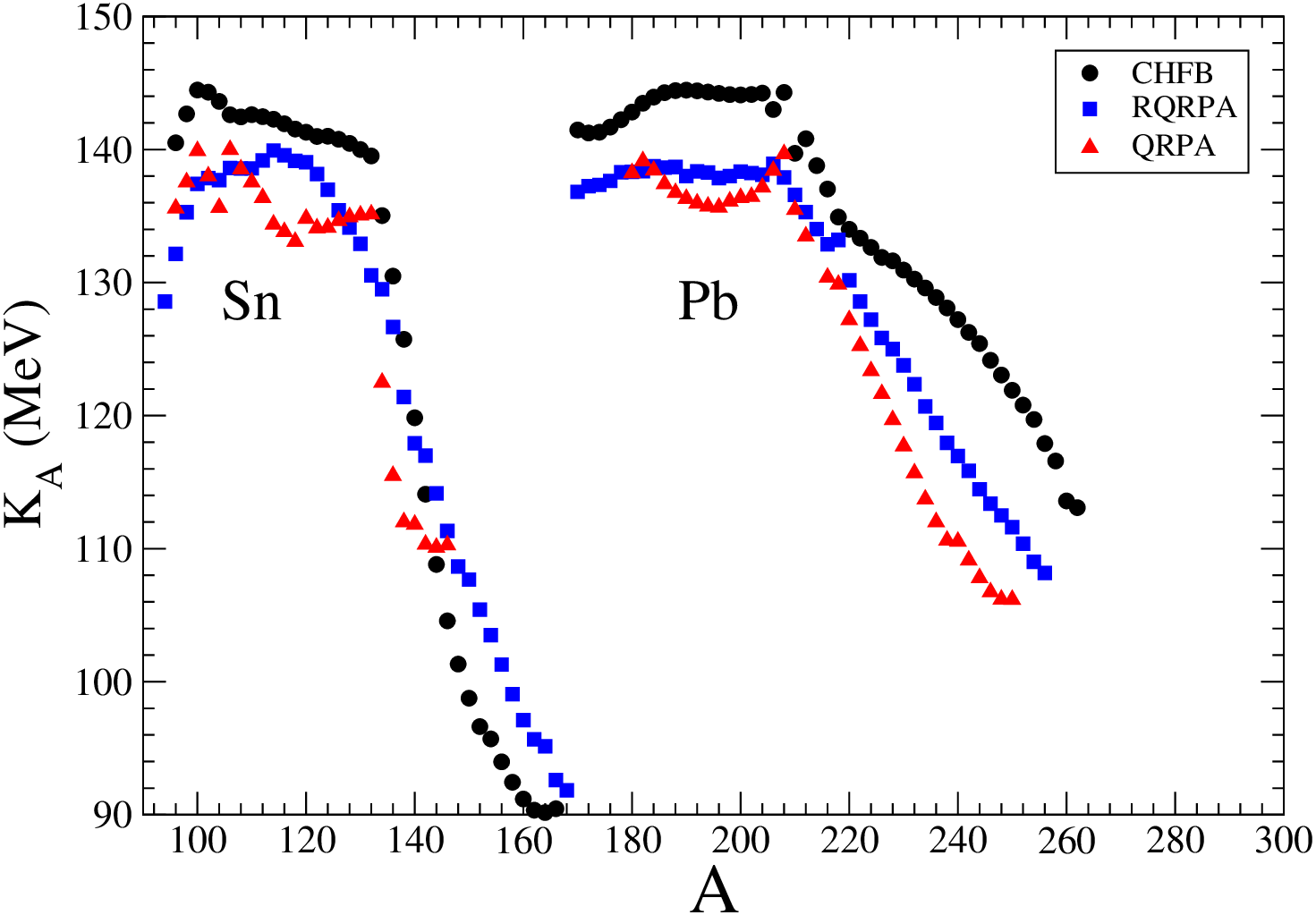}}
\caption{(Color online) Nuclear incompressibility of the Sn and Pb isotopic chains, 
calculated using the microscopic Skyrme-CHFB method (circles), the
Skyrme-QRPA (triangles), and the relativistic QRPA (squares).}
\label{fig:ka}
\end{center}
\end{figure}

The values of the nuclear incompressibility calculated using the CHFB,
non-relativistic QRPA, and relativistic QRPA for the chains of Sn and 
Pb isotopes are displayed in Fig. \ref{fig:ka}. With the GMR
excitation energies $E_{\rm GMR}$ determined by the moments of the
strength distributions (cf. Eq. (\ref{eq:egsum})), K$_A$ is calculated
from Eq.~(\ref{eq:ka}). We note that the three microscopic models
predict remarkably similar isotopic trends, with values of K$_A$ for
individual nuclei that do not differ by more than 10\%. The nuclear
incompressibility $K_A$ is rather constant for stable nuclei and 
systems with moderate neutron excess, but displays a sharp decrease 
beyond the neutron shells $N=82$ for Sn, and $N=126$ for Pb isotopes. 
This behavior of $K_A$ was already observed in Ref. \cite{ves12}, in
which the isotopic Pb and Sn chains were calculated in the QRPA with
both the SLy4 and UNEDF0 effective interactions.  The rather constant
values of $K_A$ shown in Fig. \ref{fig:ka} are close to 140 MeV, in
quantitative agreement with the HO prediction (Eq. (\ref{eq:rinf2})).
This shows that the analytical approach of section \ref{analytic} B.2
provides an accurate approximation for $K_A$ in isotopes close to
stability, for which valence neutrons typically occupy the same major
shell.  In the model of Sec. \ref{analytic}, the energy of the
monopole is taken as 2$\hbar\omega$, in accordance with the experimental
trend ($\hbar\omega\approx$ 41 $A^{-1/3}$ and E$_{\rm
ISGMR}\approx$ 80 $A^{-1/3}$). 
Many energy functionals are
characterized by values of the effective mass $m^*/m$ that are smaller
than 1. In this case it is the RPA attractive residual interaction
that lowers the monopole energy to its empirical value, even when the 
unperturbed energy is larger than 2$\hbar\omega$, that is, 
(2$\hbar\omega m/m^*$). Therefore, one understands that the HO
estimate of K$_A\simeq$140 MeV is close to the results obtained
EDF calculations. Even though the latter include effects of the 
effective mass and the residual interaction, there is an approximate 
cancellation of these effects.
%--------------------------------------------------------------

\begin{figure}[tb]
\begin{center}
\scalebox{0.35}{\includegraphics{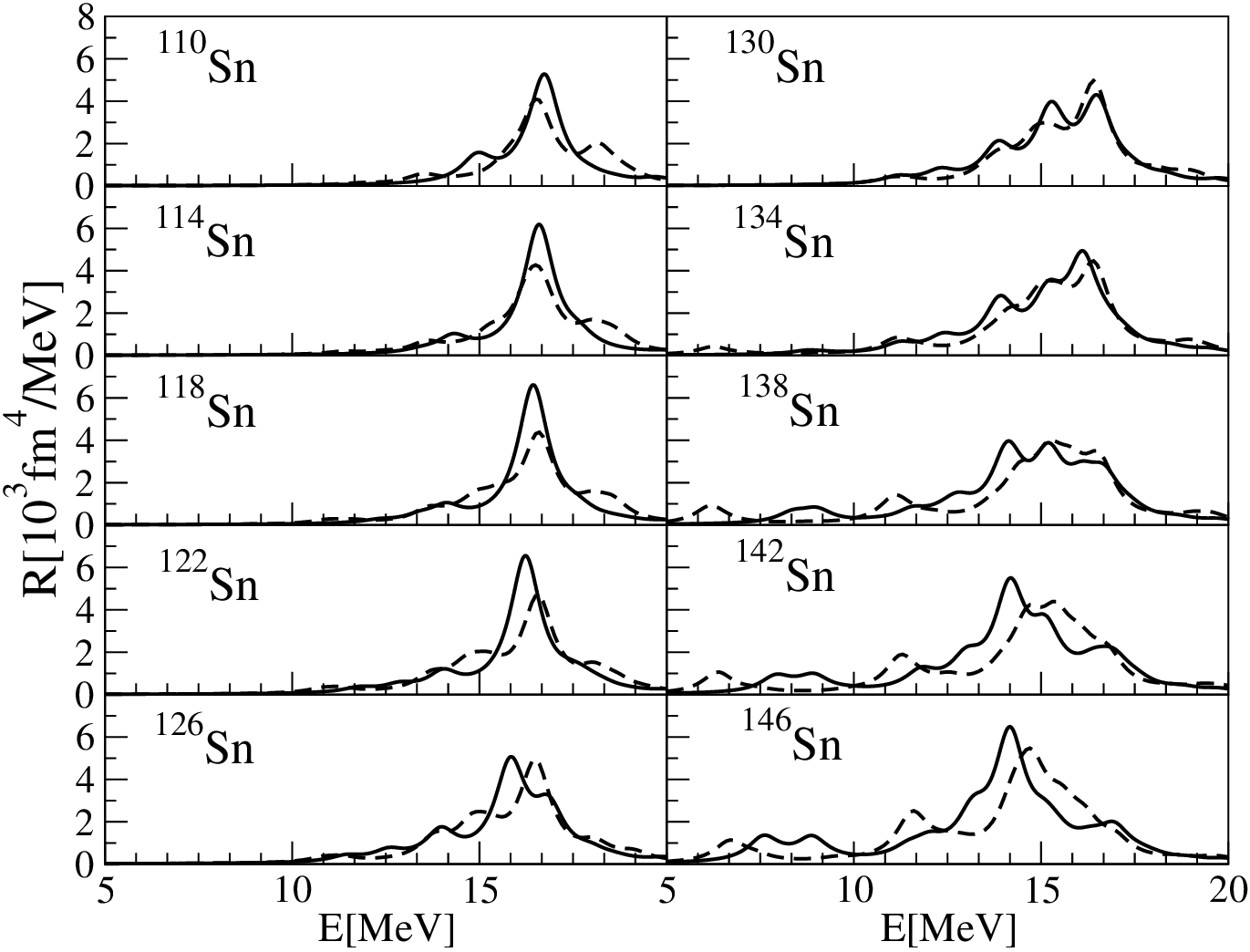}}
\caption{Isoscalar monopole response in Sn isotopes, calculated using 
the relativistic QRPA with the DD-ME2 functional (solid), 
and the QRPA with the functional SLy5 (dashed). The response is folded 
with a Lorentzian of 1 MeV width.}
\label{fig:snr}
\end{center}
\end{figure}

\begin{figure}[tb]
\begin{center}
\scalebox{0.35}{\includegraphics{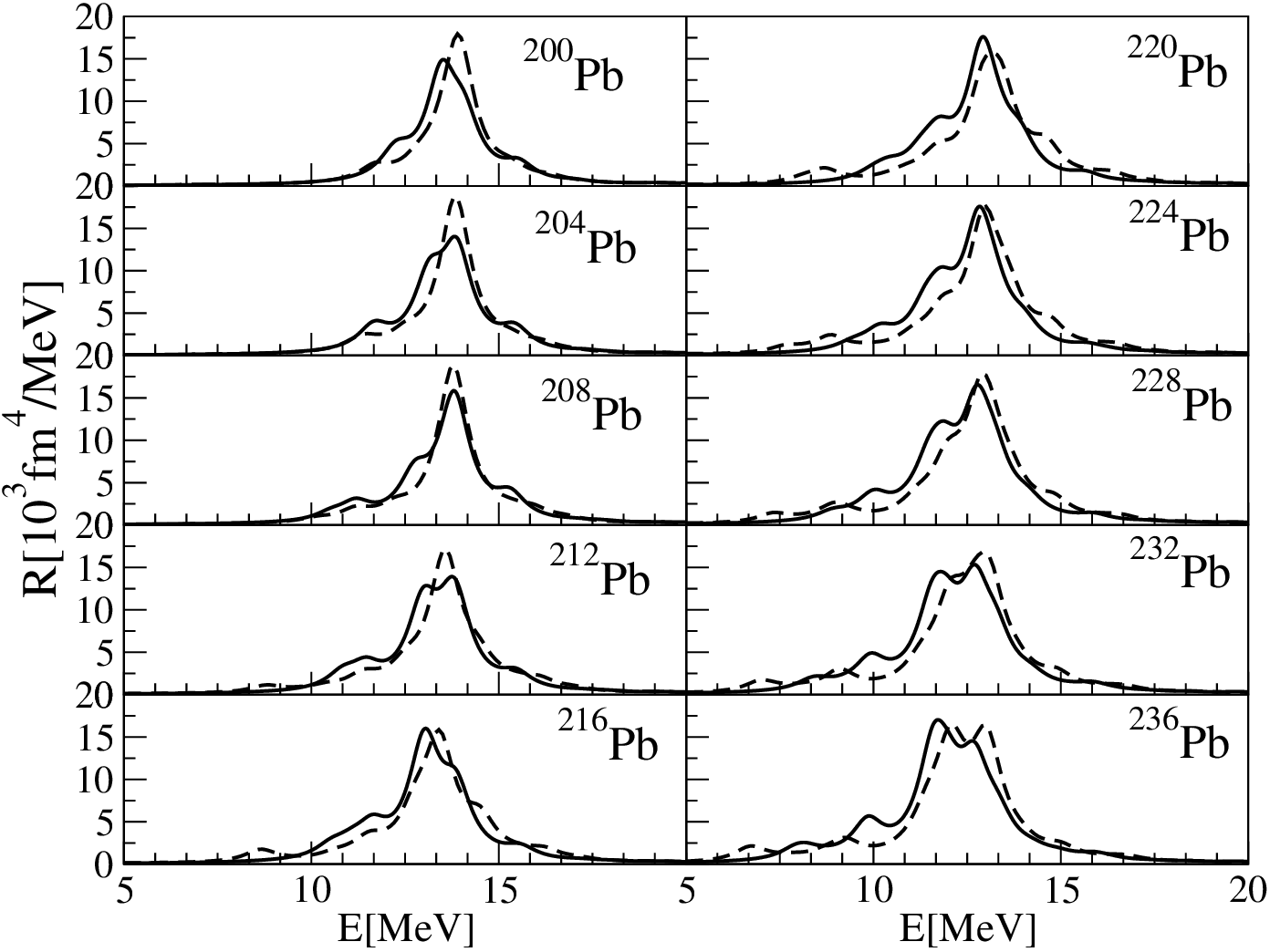}}
\caption{Same as described in the caption to Fig. \ref{fig:snr} but for the Pb isotopes.}
\label{fig:pbr}
\end{center}
\end{figure}

The sudden decrease of K$_A$ beyond the doubly magic nuclei $^{132}$Sn
and $^{208}$Pb is, of course, related to the opening of new major
shells in which monopole particle-hole configurations are formed. This
leads to the appearance of a soft monopole mode that increases the
monopole moment $m_{-1}$ and, consequently, decreases the deduced
value of K$_A$. Soft monopole modes have been predicted in several
studies \cite{ham98,cap09,kha11,yuk13}.  In Ref. \cite{cap09} the
effect of a soft monopole mode on the excitation energy of the GMR was
analyzed for Ni and Ca isotopes using the Skyrme-QRPA model, and
similar conclusions were reached. Figs. \ref{fig:snr} and
\ref{fig:pbr} display the isoscalar monopole response in a number of
Sn and Pb isotopes, calculated using both the non-relativistic and
relativistic QRPA with the functionals SLy5 and DD-ME2, respectively.
The emergence of a soft monopole mode for exotic nuclei far from
stability is clearly visible, and can be related to the decrease of
the nuclear incompressibility $K_A$ in neutron-rich nuclei. 

The steep decrease of $K_A$ shown in Fig. \ref{fig:ka} raises the question 
of the effect of magicity on the nuclear incompressibility. Despite a specific 
prediction \cite{kha09b}, no evidence of the mutual magicity enhancement (MEM) 
effect for the incompressibility was found experimentally \cite{gar12}. The absence 
of the MEM effect was further confirmed
by additional studies \cite{lig12,ves12}. However, the present
results show that just beyond major-shell closure the nuclear
incompressibility decreases because of the appearance of the soft monopole
mode. One could, therefore, conclude that a magicity
effect on the incompressibility is predicted, although it is not the MEM effect.

\begin{figure}[tb]
\begin{center}
\scalebox{0.35}{\includegraphics{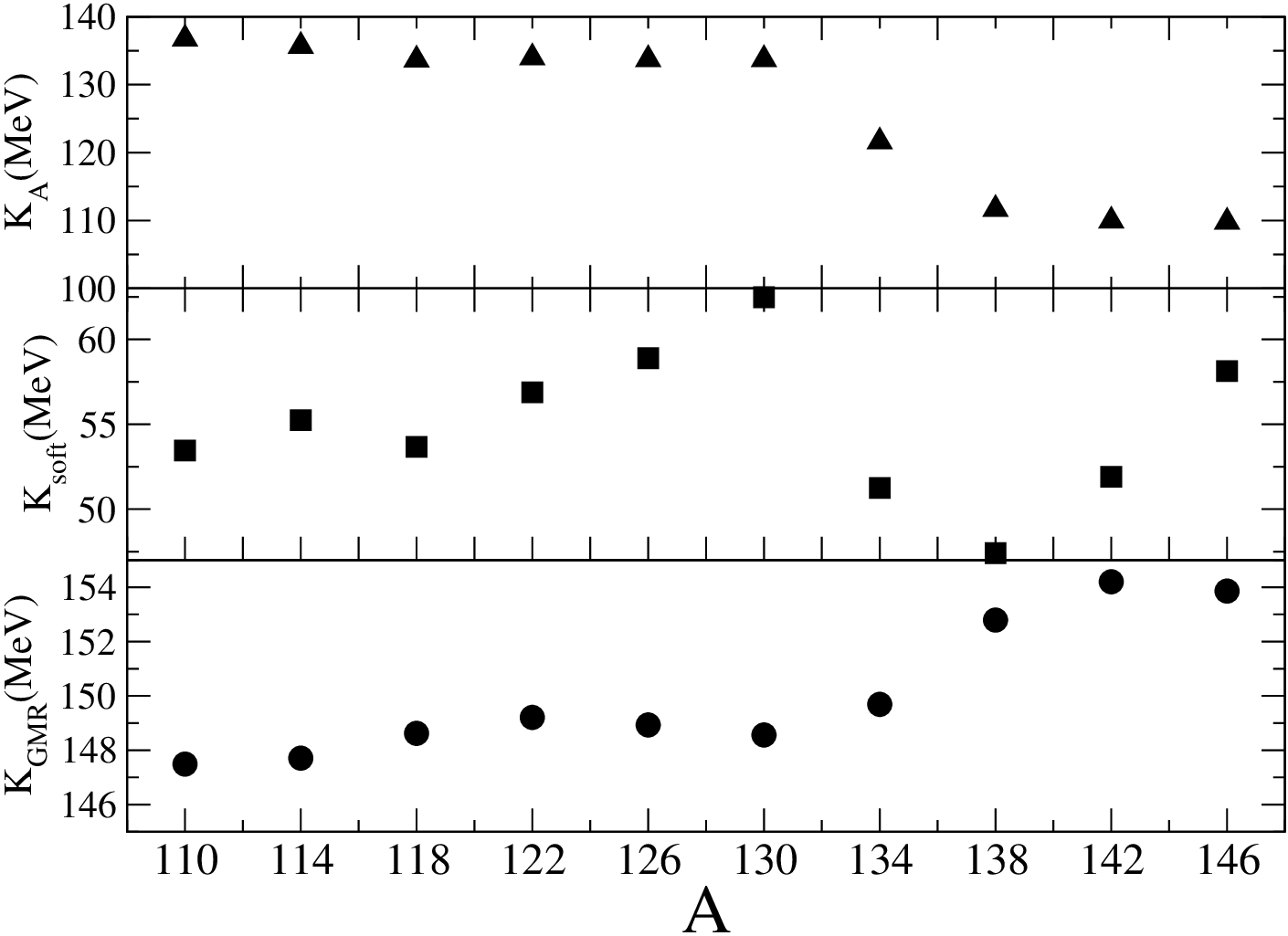}}
\caption{Nuclear incompressibility in the Sn isotopic chain 
calculated using the microscopic Skyrme-QRPA with the functional SLy5.
The total nuclear incompressibility $K_A$ is shown in the upper panel, 
whereas the values determined by the monopole response in the 
low-energy and GMR regions are plotted in the middle and lower 
panels, respectively.}
\label{fig:kasn}
\end{center}
\end{figure}

\begin{figure}[tb]
\begin{center}
\scalebox{0.35}{\includegraphics{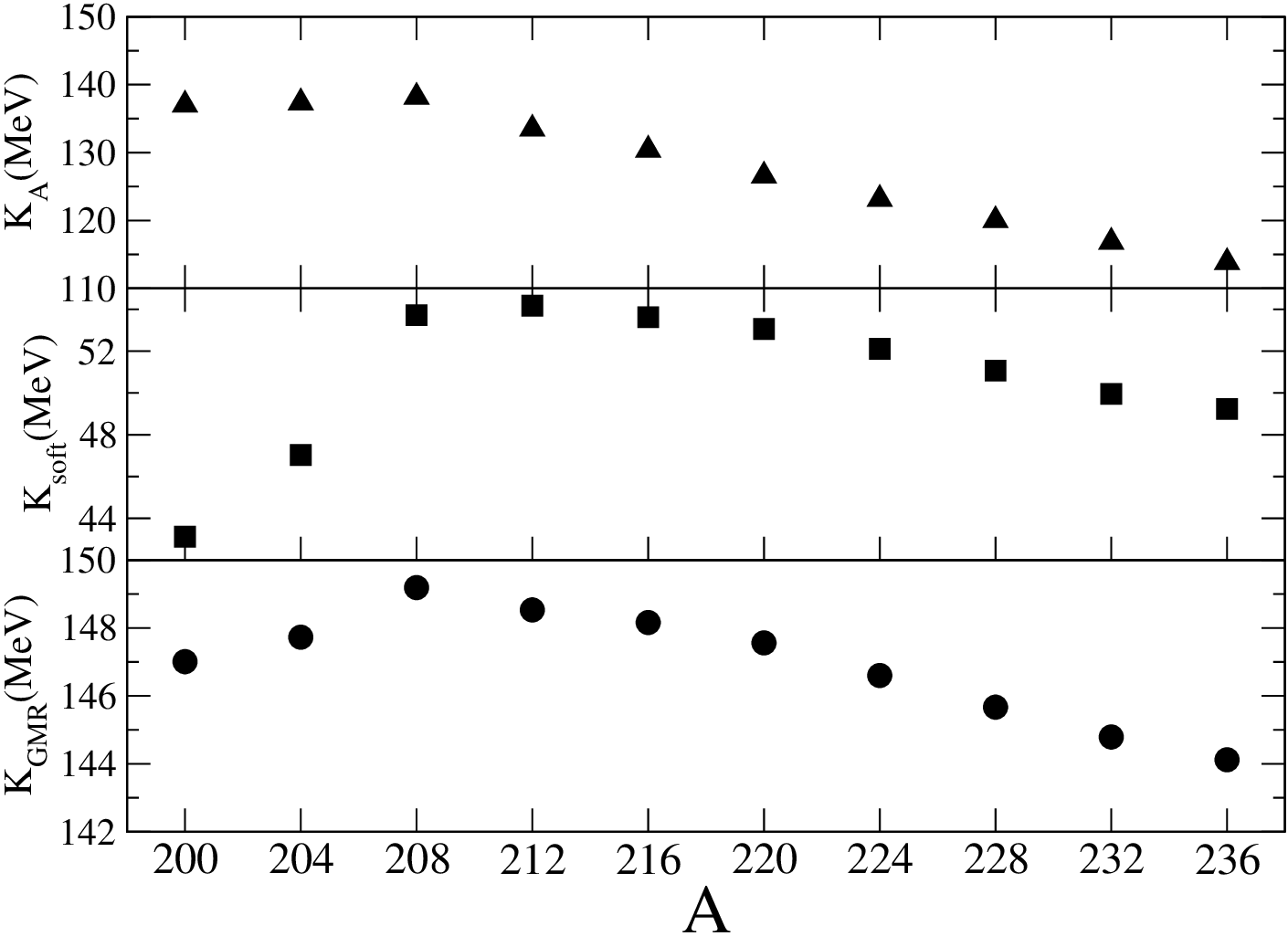}}
\caption{Same as described in the caption to Fig. \ref{fig:kasn} but for the Pb isotopes.}
\label{fig:kapb}
\end{center}
\end{figure}

Figs. \ref{fig:snr} and \ref{fig:pbr} clearly exhibit the emergence of a soft
monopole mode in the QRPA monopole response functions of  
neutron-rich nuclei. It might, therefore, be useful to calculate the 
values of the incompressibility $K_A$ separately in the low-energy 
region of the soft mode,  and in the energy region of the GMR. The 
results are shown in Figs. \ref{fig:kasn} and \ref{fig:kapb}, where the 
isotopic dependence of this values, calculated using the QRPA with 
the functional SLy5, is plotted for a series of  Sn and Pb nuclei, respectively. 
The boundary between the low-energy and GMR regions is set 
arbitrarily at 13.5 MeV for Sn
isotopes, and at 11 MeV for Pb isotopes. $K_{soft}$ ($K_{GMR}$) is
determined from the monopole response below (above) this
threshold. The isotopic dependence of $K_{soft}$ and $K_{GMR}$ 
reflects the evolution of the soft monopole mode and the GMR. 
While in the case of the Sn
isotopes no definite trend is found, for the Pb
nuclei both $K_{soft}$ and $K_{GMR}$ show a steady decrease with
increasing neutron number. 

\section{Conclusions  \label{conclusion}}

The incompressibility of finite fermionic systems has been studied
using several analytical and numerical approaches.  Analytic relations
obtained using a simple free Fermi gas or harmonic oscillator (HO)
models point to a direct link between the incompressibility of the
system and the zero point kinetic energy of constituent particles.
The incompressibility is thus related to the localization of the
individual  fermions. For atomic nuclei the HO model predicts 
the incompressibility  $K_A\approx$ 5$T_0 \approx 140$ MeV.

To quantitatively analyze the evolution of nuclear incompressibility in 
isotopic chains of medium-heavy and heavy nuclei, microscopic EDF-based 
approaches have been used: the 
constrained HFB method and the QRPA based on both non-relativistic and 
relativistic density functionals. All microscopic models predict a remarkably 
similar trend for the incompressibility: an almost constant behavior in stable 
nuclei and systems not far from stability, with values of $K_A$ close to 
those predicted by the HO model, and a steep decrease of $K_A$ in 
nuclei with pronounced neutron excess, caused by the emergence of 
a soft monopole mode in neutron-rich nuclei. It would be important 
to have an experimental confirmation of the soft monopole
mode in neutron-rich nuclei and,  more generally, to study 
the behavior of the monopole response in exotic nuclei and/or beyond 
shell closures. For instance, measurement of the GMR beyond 
$^{132}$Sn and $^{208}$Pb would be very interesting.

\section{Acknowledgments}

This work was supported by the Institut Universitaire de France, 
and  by the MZOS - project 1191005-1010.


\begin{thebibliography}{99}
\bibitem{tli07} T. Li, U. Garg, Y. Liu, R. Marks, B.K. Nayak,
P.V. Madhusudhana Rao, M. Fujiwara, H. Hashimoto, K. Kawase, K.
Nakanishi, S. Okumura, M. Yosoi, M. Itoh, M. Ichikawa, R. Matsuo, T.
Terazono, M. Uchida, T. Kawabata, H. Akimune, Y. Iwao, T. Murakami, H.
Sakaguchi, S. Terashima, Y. Yasuda, J. Zenihiro, and M. N. Harakeh, 
Phys. Rev. Lett. 99, 162503 (2007).
\bibitem{tli10} T. Li, U. Garg, Y. Liu, R. Marks, B.K. Nayak, P.V.
Madhusudhana Rao, M. Fujiwara, H. Hashimoto, K. Nakanishi, S. Okumura, M. Yosoi, M. Ichikawa, M. Itoh, R. Matsuo, T. Terazono, M. Uchida, Y. Iwao, T. Kawabata, T. Murakami, H. Sakaguchi, S. Terashima, Y. Yasuda,
J. Zenihiro, H. Akimune, K. Kawase, M.N. Harakeh, Phys. Rev. 
C81, 034309 (2010).
\bibitem{gar09} D. Patel, U. Garg, M. Fujiwara, H. Akimune, G.P.A.
Berg, M.N. Harakeh, M. Itoh, T. Kawabata, K. Kawase, B.K. Nayak, T.
Ohta, H. Ouchi, J. Piekarewicz, M. Uchida, H.P. Yoshida, M. Yosoi,
Phys. Lett. B718, 447 (2012).
\bibitem{jli08} J. Li, G. Col\`o and J. Meng, Phys. Rev. C78, 
064304 (2008).
\bibitem{kha09} E. Khan, Phys. Rev. C80, 011307(R) (2009)
\bibitem{ves12} P. Vesel\'y, J. Toivanen, B.G. Carlsson, J.
Dobaczewski, N. Michel, and A. Pastore, Phys. Rev. C86, 024303 (2012).
\bibitem{sor08} O. Sorlin and M.-G. Porquet, Progress in Particle and Nuclear Physics 61, 602 (2008).
\bibitem{kha12} E. Khan, J. Margueron and I. Vida\~na, Phys. Rev. Lett. 109, 092501 (2012).
\bibitem{mon08} C. Monrozeau, E. Khan, Y. Blumenfeld, C.E. Demonchy, W.
Mittig, P. Roussel-Chomaz, D. Beaumel, M. Caama\~no, D. Cortina-Gil, J.
P. Ebran, N. Frascaria, U. Garg, M. Gelin, A. Gillibert, D. Gupta, N.
Keeley, F. Mar\'echal, A. Obertelli, and J-A. Scarpaci, 
Phys. Rev. Lett. 100, 042501 (2008).
\bibitem{van10} M. Vandebrouck et al., Exp. E456a performed at GANIL
(2010).
\bibitem{kha09b} E. Khan, Phys. Rev. C80, 057302 (2009)
\bibitem{lig12} Li-Gang Cao, H. Sagawa, and G. Col\`o, Phys. Rev. C 86, 054313 (2012)
\bibitem{gar12} U. Garg (Private communication).
\bibitem{sch06} S. Shlomo, V.M. Kolomietz, and G. Col\`o, Eur. Phys. J. A 30, 23 (2006). 
\bibitem{ben03} M. Bender, P.-H. Heenen, P.-G. Reinhard, Rev. Mod. Phys. 75,  121-180 (2003).
\bibitem{vre05} D. Vretenar, A.V. Afanasjev, G.A. Lalazissis, P. Ring, Phys. Rep. 409, 101-259 (2005).
\bibitem{bla80} J.-P. Blaizot, Phys. Rep. 64, 171 (1980).
\bibitem{boh79} O. Bohigas, A.M. Lane and J. Martorell, Phys. Rep. 51, 267 (1979).
\bibitem{ham98} I. Hamamoto, H. Sagawa, and X. Z. Zhang, Phys. Rev. C53, 765 (1996).
\bibitem{cap09} L. Capelli, G. Col\`o and J. Li, Phys. Rev. C79, 054329 (2009).
\bibitem{kha11} E. Khan, N. Paar, and D. Vretenar, Phys. Rev. C84, 051301(R) (2011).
\bibitem{lan67} L. Landau and E. Lifchitz, Statistical Physics, MIR Edition, Moscow (1967).
\bibitem{tho61} D.J. Thouless, Nucl. Phys. 22, 78 (1961).
\bibitem{fet71} A.L. Fetter and J.D. Walecka, {\it Quantum Theory of Many-Particle 
Systems} (McGraw-Hill, New York, 1971).
\bibitem{mot96} B. Mottelson, Nuclear Structure, Les Houches, Session LXVI, 25 (1996).
\bibitem{VdW} M. Harakeh, A. Van der Woude, {\it Giant Resonances}, Oxford University Press (2001).
\bibitem{paa07} N. Paar, D. Vretenar, E. Khan and G. Col\`o, Rep. Prog. Phys. 70, 691 (2007).
\bibitem{cha98} E. Chabanat, P. Bonche, P. Haensel, J. Meyer, R. Schaeffer, Nucl. Phys. A635, 231
(1998).
\bibitem{kha09c} E. Khan, M. Grasso, J. Margueron, Phys. Rev. C80, 044328 (2009).
\bibitem{rin80} P. Ring, P. Schuck, The Nuclear Many-Body Problem, Springer-Verlag, Heidelberg, (1980).
\bibitem{colo13} G. Col\`o, L.G. Cao, N. Van Giai, L. Capelli, Comp. Phys. Comm. 184, 142 (2013).
\bibitem {VALR.05} D. Vretenar, A. V. Afanasjev, G. A. Lalazissis and P. Ring, Phys. Rep. 409, 101 (2005).
\bibitem{NVFR.02} T. Nik\v si\' c, D. Vretenar, P. Finelli, and P. Ring, Phys. Rev. C 66, 024306 (2002).
\bibitem{PRNV.03} N. Paar, P. Ring, T. Nik\v si\' c and D. Vretenar, Phys. Rev. C 67, 034312 (2003).
\bibitem{BGG.91} J. F. Berger, M. Girod and D. Gogny, Comp. Phys. Comm. 63, 365 (1991).
\bibitem{LNVR.05} G. A. Lalazissis, T. Nik{\v{s}}i{\'{c}}, D. Vretenar
and P. Ring, Phys. Rev. C 71 024312 (2005).
\bibitem{yuk13} E. Y\"{u}ksel, E. Khan, K. Bozkurt, in preparation.
\end{thebibliography}
\end{document}